%% file: main.tex
\documentclass[runningheads]{llncs}
\usepackage[T1]{fontenc}
\usepackage{graphicx,verbatim}
\usepackage{multicol, xcolor, colortbl}
\usepackage{hyperref}
\usepackage{pdfpages}
\usepackage{censor}
\usepackage{booktabs} 
\usepackage{tabularx}
\usepackage{array}
\usepackage{orcidlink}
\usepackage{placeins}
\usepackage{amsmath}
\usepackage{bm}

\begin{document}
\title{PyPeT: A Python Perfusion Tool for Automated Quantitative Brain CT and MR Perfusion Analysis}

%
\begin{comment}  %% Removed for anonymized MICCAI 2025 submission
\author{First Author\inst{1}\orcidID{0000-1111-2222-3333} \and
Second Author\inst{2,3}\orcidID{1111-2222-3333-4444} \and
Third Author\inst{3}\orcidID{2222--3333-4444-5555}}
%
\authorrunning{F. Author et al.}
\institute{Princeton University, Princeton NJ 08544, USA \and
Springer Heidelberg, Tiergartenstr. 17, 69121 Heidelberg, Germany
\email{lncs@springer.com}\\
\url{http://www.springer.com/gp/computer-science/lncs} \and
ABC Institute, Rupert-Karls-University Heidelberg, Heidelberg, Germany\\
\email{\{abc,lncs\}@uni-heidelberg.de}}

\end{comment}

% \author{Anonymized Authors}  %% Added for anonymized MICCAI 2025 submission
% \authorrunning{Anonymized Author et al.}
% \institute{Anonymized Affiliations \\
%     \email{email@anonymized.com}}

\author{Marijn Borghouts\inst{1}~\orcidlink{0009-0002-3820-3957} \and Ruisheng Su\inst{1}~\orcidlink{0000-0002-5013-1370}}
\index{Borghouts, Marijn; }
\authorrunning{M. Borghouts et al.} 
\titlerunning{PyPeT: Python Perfusion Tool for CTP and MRP}
\institute{
    $^1$Department of Biomedical Engineering, Eindhoven University of Technology, Eindhoven, The Netherlands \\
    Correspondence: \email{m.m.borghouts@tue.nl}
}

\maketitle
\begin{abstract}
% The abstract should briefly summarize the contents of the paper in 150--250 words.  If you are to include a link to your Repository, please make sure it is anonymized for the double-blind review phase.
\input{text/0.Abstract}

% Authors must provide keywords and are not allowed to remove this Keyword section.

\end{abstract}

\section{Introduction}
\input{text/1.Introduction}

\section{Methodology}
\input{text/2.Methodology}

\FloatBarrier 
\section{Validation Experiments}
\input{text/3.Experiments}

\FloatBarrier 
\section{Results}
\input{text/4.Results}

\FloatBarrier
\section{Discussion and Conclusion}

\input{text/5.Conclusion_and_Discussion}

\begin{comment}  %% removed for anonymized MICCAI 2025 submission.
    
    % The following acknowledgement and disclaimer sections should be removed for the double-blind review process.  
    % If and when your paper is accepted, reinsert the acknowledgement and the disclaimer clause in your final camera-ready version.

\begin{credits}
\subsubsection{\ackname} A bold run-in heading in small font size at the end of the paper is
used for general acknowledgments, for example: This study was funded
by X (grant number Y).

\subsubsection{\discintname}
It is now necessary to declare any competing interests or to specifically
state that the authors have no competing interests. Please place the
statement with a bold run-in heading in small font size beneath the
(optional) acknowledgments\footnote{If EquinOCS, our proceedings submission
system, is used, then the disclaimer can be provided directly in the system.},
for example: The authors have no competing interests to declare that are
relevant to the content of this article. Or: Author A has received research
grants from Company W. Author B has received a speaker honorarium from
Company X and owns stock in Company Y. Author C is a member of committee Z.
\end{credits}

\end{comment}
\FloatBarrier % to prevent images from being placed after the reference
\bibliographystyle{splncs04} 
\bibliography{export}

\newpage
\section{Appendices}
\input{text/6.Appendices}

\end{document}

%% file: text/0.Abstract.tex
Computed tomography perfusion (CTP) and magnetic resonance perfusion (MRP) are widely used in acute ischemic stroke assessment and other cerebrovascular conditions to generate quantitative maps of cerebral hemodynamics. While commercial perfusion analysis software exists, it is often costly, closed source, and lacks customizability. This work introduces PyPeT, an openly available \textbf{Py}thon \textbf{Pe}rfusion \textbf{T}ool for head CTP and MRP processing. PyPeT is capable of producing cerebral blood flow (CBF), cerebral blood volume (CBV), mean transit time (MTT), time-to-peak (TTP), and time-to-maximum (Tmax) maps from raw four-dimensional perfusion data. 
PyPeT aims to make perfusion research as accessible and customizable as possible. This is achieved through a unified framework in which both CTP and MRP data can be processed, with a strong focus on modularity, low computational burden, and significant inline documentation. PyPeT's outputs can be validated through an extensive debug mode in which every step of the process is visualized. Additional validation was performed via visual and quantitative comparison with reference perfusion maps generated by three FDA-approved commercial perfusion tools and a research tool. These comparisons show a mean SSIM around 0.8 for all comparisons, indicating a good and stable correlation with FDA-approved tools. The code for PyPeT is openly available at our GitHub \url{https://github.com/Marijn311/CT-and-MR-Perfusion-Tool}

%% file: text/1.Introduction.tex
\subsection{Background}
\begin{figure}[ht]
\centering
\includegraphics[width=\textwidth]{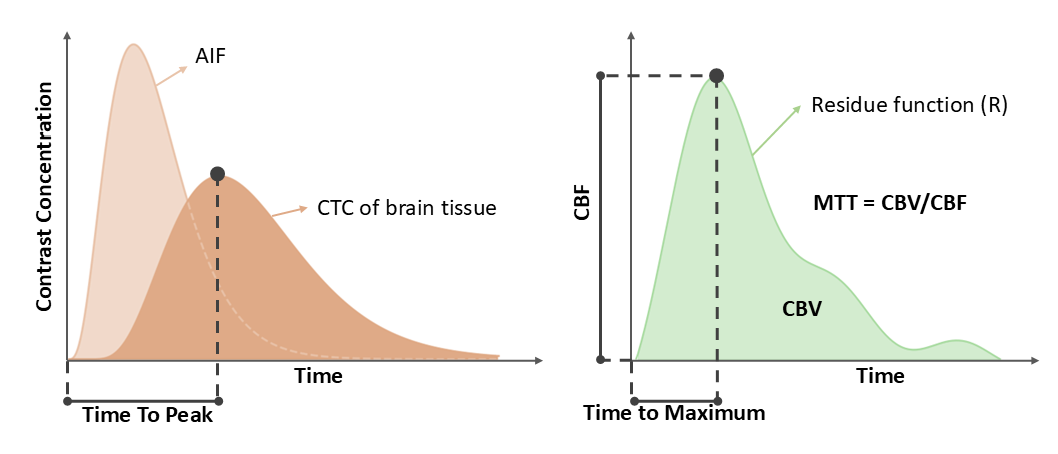}
\caption{Contrast time curves (CTC), residue function, and the derived perfusion parameters.}
\label{fig:perfusion_definitions}
\end{figure}

Perfusion imaging is most commonly employed in the acute management of ischemic stroke, but it also has applications in other cerebrovascular conditions, including vasospasms, brain tumors, and head trauma. CTP and MRP are perfusion imaging techniques in which a contrast agent is injected into the cerebral circulation, while the brain is continuously imaged using CT or MR as the contrast agent passes through. By isolating the contrast time curves (CTC) and deconvolving them with the arterial input function (AIF), one can obtain the residue function (R). From this residue function, one can extract the CBF, CBV, MTT, and Tmax to visualize the cerebral hemodynamics, see Figure~\ref{fig:perfusion_definitions} and Equations 1-6. Obtaining stable perfusion maps from raw perfusion images, however, require a significant amount of processing steps.   

\begin{align}
R(t) &= \frac{1}{\mathrm{CBF}} \cdot \left( \mathrm{CTC}(t) \oslash \mathrm{AIF}(t) \right) \\
\mathrm{TTP} &= \arg\max_t \left(\mathrm{AIF}(t)\right) \\
\mathrm{CBF} &= \max_t R(t) \\
\mathrm{CBV} &= \int R(t)\, dt \\
\mathrm{Tmax} &= \arg\max_t R(t) \\
\mathrm{MTT} &= \frac{\mathrm{CBV}}{\mathrm{CBF}}
\end{align}

\paragraph{Variable definitions.}
\begin{description}
    \item[CTC$(t)$] Concentration time curve, representing the measured contrast concentration in the tissue as a function of time.
    \item[AIF$(t)$] Arterial input function, describing the CTC in an upstream artery.
    \item[$R(t)$] Residue function, denoting the fraction of contrast agent remaining in the tissue at time $t$ following an idealized impulse injection.
    \item[CBF] Cerebral blood flow, defined as the maximum value of the residue function and expressing the volume of blood delivered per unit tissue per unit time.
    \item[CBV] Cerebral blood volume, computed as the integral of the residue function and representing the volume of blood contained in a unit mass of tissue.
    \item[TTP] Time to peak of the arterial input function, indicating the arrival time of contrast in the supplying artery.
    \item[Tmax] Time to the maximum of the residue function, indicating the arrival time of contrast in the tissue.
    \item[MTT] Mean transit time, defined as $\mathrm{CBV} / \mathrm{CBF}$ and reflecting the average time blood spends within the tissue.
\end{description}

\subsection{Problem Statement}
Several FDA-approved commercial software solutions are available for perfusion image processing (e.g., Icobrain CVA, Viz CTP, iSchemaView RAPID, Canon MR-Clinical-Suite \cite{Chandrabhatla2023}). While these tools are effective, they are often expensive, closed-source, and limited in flexibility, making them less suitable for research. Consequently, there is a need for open alternatives that allow for easy customization and transparency.

\subsection{Related Works}
Some open approaches to perfusion image processing have been developed. For CTP, examples include the codebases described in \cite{Turtleizzy,Lirette2018}, which both implement a singular value decomposition (SVD)-based deconvolution method. Another contribution is the work of \cite{Bennink2016}, who proposed a box nonlinear regression technique to improve upon SVD-based deconvolution for generating perfusion maps. More recently, machine learning approaches have been explored for CTP processing. For instance, \cite{Vries2023} employed physics-informed neural networks, while \cite{Gava2023} applied a U-Net architecture to estimate perfusion maps directly from raw CTP data.

For MRP, open approaches include the works of \cite{Rodicio2023,Peruzzo2011,Gordaliza2015} who developed SVD-based deconvolution implementations. Additionally, there is the work by~\cite{Levy2021} who adapted the work of~\cite{Peruzzo2011}. There is also the project of~\cite{Houdt2024} which aims to collect and validate open-source code fragments for MRP processing. More recently, there have been machine learning based approaches for MRP processing as well, such as the work by \cite{Lohrke2024}, which uses GAN to generate perfusion maps.

Although these open approaches are all valuable contributions to the field, and this list is surely not all-encompassing, each comes with its own limitations. Some represent only basic implementations and therefore lack features expected from a comprehensive perfusion analysis toolbox. Others provide limited or insufficient documentation, reducing their usability and customizability. Certain tools demand substantial computational resources, requiring high-performance CPUs or GPUs to function effectively. Some tools lack proper validation strategies. Moreover, all of these solutions are restricted to a single imaging modality. As a result, there remains a need for further progress in the development of open perfusion processing tools.

Therefore, PyPeT aims to contribute the first open and unified framework capable of processing both CTP and MRP data within a single tool. The software includes extensive inline documentation and is accompanied by this manuscript, which explains its design and functionality in detail. In addition, PyPeT offers a comprehensive debug mode that enables early detection of potential processing errors and includes validation against multiple FDA-approved perfusion analysis toolboxes.

\subsection{Design Philosophy}
The primary objective of this work is to make perfusion image processing as accessible as possible for researchers. To this end, we introduce PyPeT, an open \textbf{Py}thon \textbf{Pe}rfusion \textbf{T}ool. Python was chosen as the implementation language because of its widespread usage, clean syntax, expressiveness, and ecosystem of libraries~\cite{Ekmekci2016}. Furthermore, Python is freely available, in contrast to commercial alternatives such as MATLAB. PyPeT employs an SVD-based deconvolution method to generate perfusion maps. Unlike deep learning approaches, SVD-based methods do not require training datasets or expensive hardware. The entire PyPeT pipeline can be executed on a mobility laptop in seconds. PyPeT's codebase was deliberately structured in a modular fashion with extensive comments to support flexibility, customization, and future extensions by the research community. Finally, PyPeT introduces a unified framework for handling both CTP and MRP data. This dual compatibility broadens its applicability across different modalities and research contexts, making it a versatile tool for perfusion imaging studies. This report provides both a technical overview of the perfusion processing pipeline and practical guidance for researchers aiming to customize or validate its outputs.

%% file: text/2.Methodology.tex
\subsection{Perfusion Processing Steps}
PyPeT's implementation was loosely based upon the publicly available codebases of \cite{Turtleizzy,Peruzzo2011}, which provide pipelines for head CT and MR perfusion processing, respectively. The main steps of the PyPeT pipeline are described stepwise in the following sections.

First, a raw four-dimensional perfusion image is loaded from a \texttt{.nii.gz} file and preprocessed. The volumes are reoriented to a standard RAS+ orientation. Next, the volumes get downsampled to the desired shape if they exceed a user-defined threshold, to speed up computation. After which all 3D volumes are registered to the first volume using ANTsPy's QuickRigid method, to correct for patient motion. The volumes are then smoothed using a Gaussian filter. This smoothing step helps stabilize the results by mitigating noise in temporal measurements. Such noise can arise from small residual patient movements that may remain after motion correction. 

Next, a brain mask is generated. In case of CTP images, a combination of intensity thresholding and a fast-marching algorithm is applied to grow the mask from the largest connected components, which should be the skull. For MRP images, an intensity thresholding and active contour approach is applied. Alternatively, users may supply a precomputed brain mask with a tool of their choice. 

The onset of the contrast bolus is determined by tracking the mean signal intensity over time. To ensure consistent absolute values, the mean signal is normalized to the first time point. Bolus arrival is then identified using a sliding window of three time points: when the mean normalized signal within the window changes by more than a user-defined threshold (default $5\%$) compared to the preceding window, the bolus onset is marked. The contrast baseline image ($S_{0}$) is calculated by averaging all volumes acquired prior to bolus arrival. Contrast time curves (CTCs) are derived differently for CTP and MRP. In CTP, the CTC is obtained as $\text{CTC} = S - S_{0}$, reflecting the approximately linear relationship between iodine concentration and x-ray attenuation, which directly affects signal intensity. In MRP, the CTC is computed as $\text{CTC} = \tfrac{1}{\text{echo\_time}} \cdot \log\left(\tfrac{S}{S_{0}}\right)$, since the contrast agent influences relaxation times rather than directly altering signal intensity. TTP maps mark the time at which each CTC reaches its peak, with t=0 set to the contrast bolus onset.

The arterial input function (AIF) is extracted using an adaptive thresholding approach. First, the distribution of the TTP and the area under the CTC are computed. A series of progressively relaxed distribution-percentile-based thresholds for TTP and AUC are then applied to identify candidate voxels likely to represent an arterial signal. Arterial voxels are expected to exhibit low TTP and high AUC. Connected component analysis is used to isolate distinct clusters of candidate AIF voxels. For each cluster, the mean CTC is extracted. Clusters are excluded if their mean CTC is excessively noisy or if the cluster size exceeds user-defined thresholds. The mean CTC of the remaining clusters is then fitted with a gamma variate function. Finally, a scoring function that combines cluster volume, curve peak, and mean fitting error is applied to select the final AIF segmentation.

After that, the residue function (R) is determined by deconvolving the CTC with the AIF. In practice, this is done via singular value decomposition (SVD). PyPeT offers an implementation for standard SVD, block-circulant SVD with Simpson's rule integration \cite{Zanderigo2009}, block-circulant SVD with manual matrix construction, and oscillation index SVD with adaptive regularization. Next, CBF is determined as the maximum of R. Tmax is the time index corresponding to this maximum. CBV is defined as the integral of R. Lastly, MTT is calculated by dividing CBV by CBF. This information is summarized in Figure~\ref{fig:perfusion_definitions} and Equations 1-6.

\subsection{Debug Mode for Intermediate Visualizations}  
PyPeT provides a debugging mode that can be enabled by setting the \texttt{DEBUG} flag in \texttt{main.py} to \texttt{True}. When activated, the pipeline generates intermediate visualizations at key processing stages. These figures allow users to verify that each step produces the expected output, making it easier to identify and correct errors before they propagate to the final results.

\begin{figure}[ht]
\makebox[\textwidth][c]{%
    \includegraphics[clip, trim=4cm 0cm 4cm 0cm, width=\textwidth]{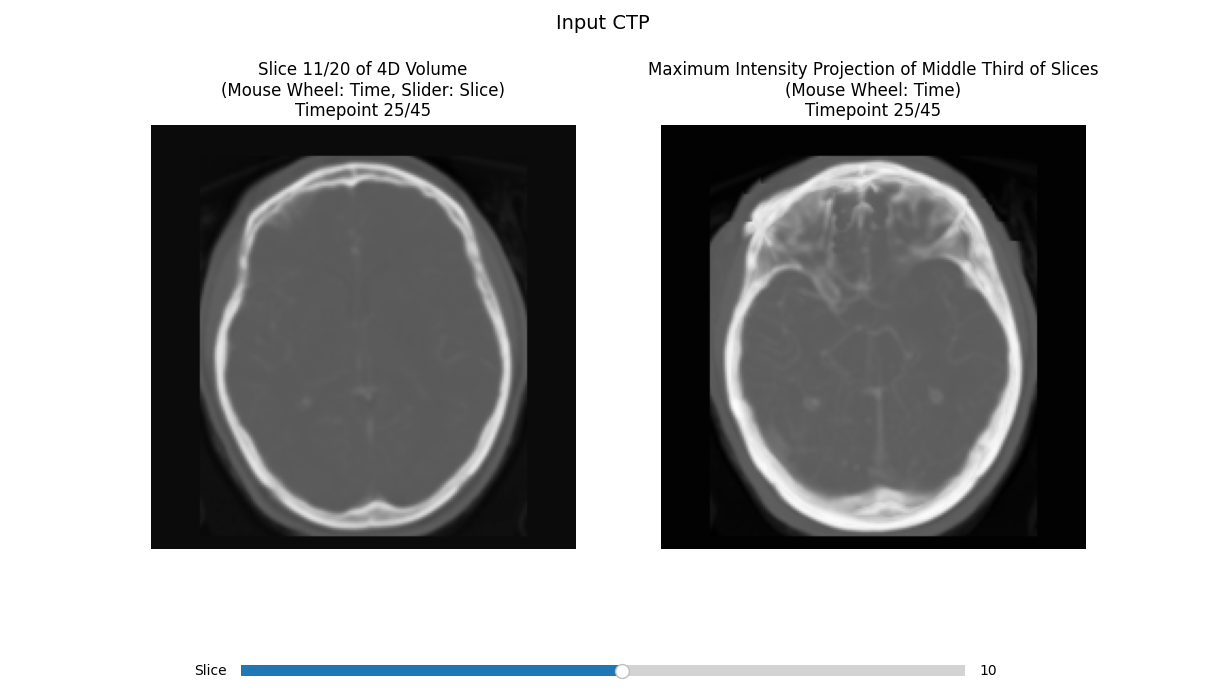}%
}
\caption{Left: snapshot of the debug viewer showing a raw 4D input CTP image. Right: corresponding maximum intensity projection of the middle third of slices.}
\label{fig:viewer_input}
\end{figure}

\begin{figure}[ht]
\makebox[\textwidth][c]{%
    \includegraphics[clip, trim=4cm 0cm 3.2cm 0cm, width=\textwidth]{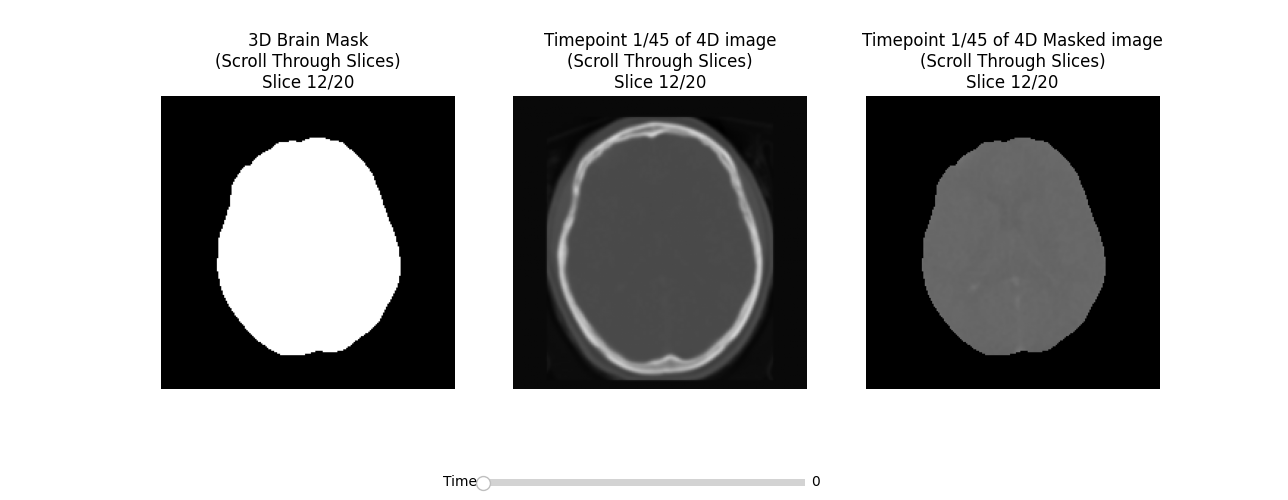}%
}
\caption{Left: snapshot of the debug viewer showing the generated mask. Middle: the corresponding perfusion image. Left: The corresponding brain segmentation.}
\label{fig:viewer_mask}
\end{figure}

\begin{figure}[ht]
\makebox[\textwidth][c]{%
    \includegraphics[width=\textwidth]{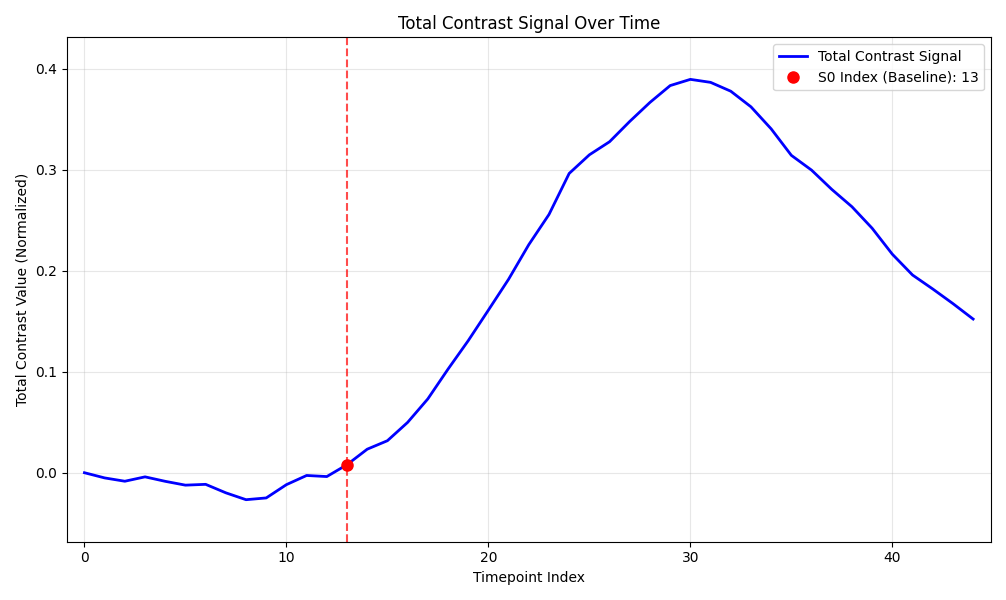}%
}
\caption{Snapshot of the debug viewer showing normalized mean signal intensity over time, with the start of the bolus highlighted.}
\label{fig:viewer_bolus}
\end{figure}

\begin{figure}[ht]
\makebox[\textwidth][c]{%
    \includegraphics[clip, trim=4cm 0cm 2.7cm 0cm, width=\textwidth]{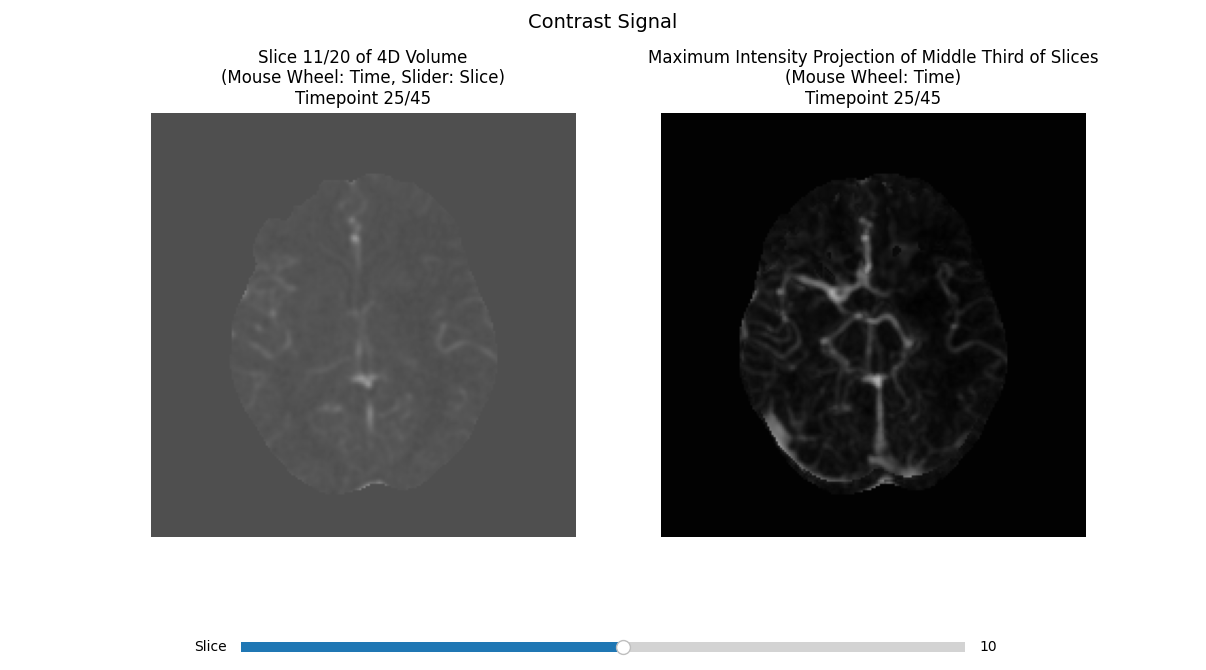}%
}
\caption{Left: snapshot of the debug viewer showing the CTC image. Right: corresponding maximum intensity projection of the middle third of slices.}
\label{fig:viewer_ctc}
\end{figure}

\begin{figure}[ht]
\makebox[\textwidth][c]{%
    \includegraphics[clip, trim=6cm 0cm 4.5cm 0cm, width=\textwidth]{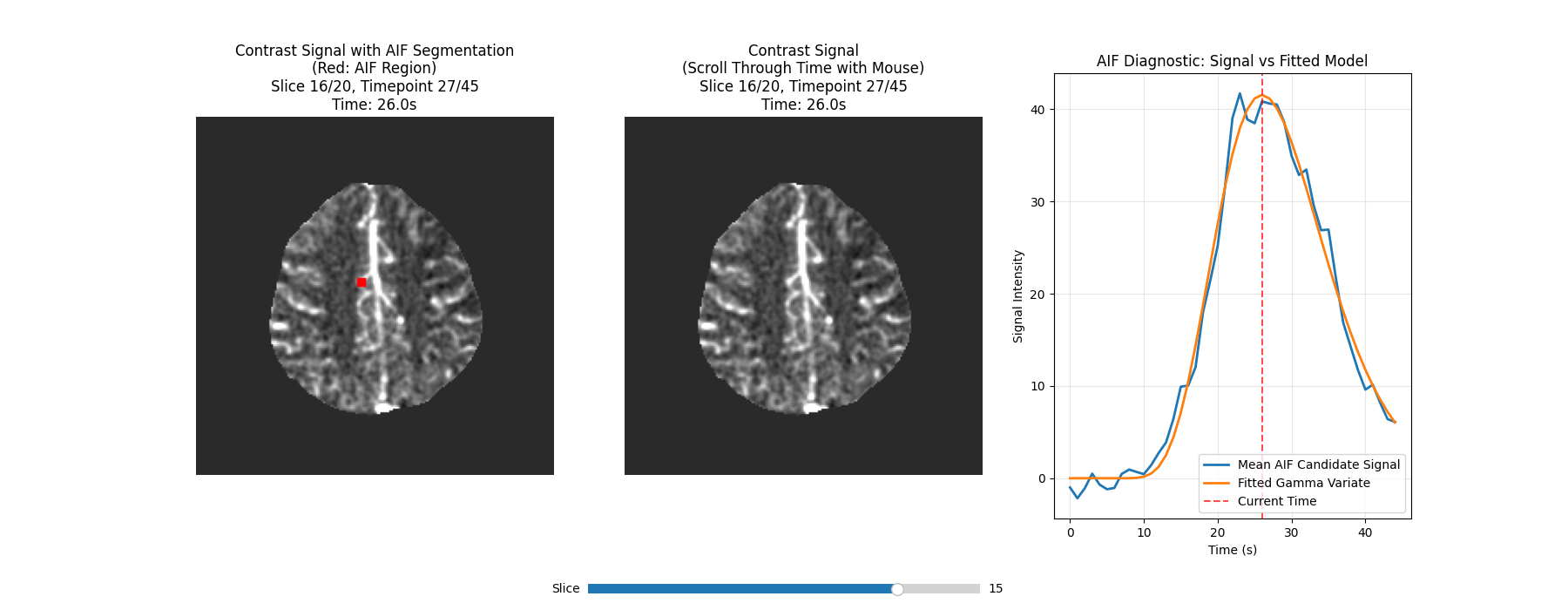}%
}
\caption{Left: slice with the AIF segmentation overlay on the CTC image. Middle: the CTC images without the AIF overlay. Right: the mean signal in the AIF segmentation, and its gamma-variate fit.}
\label{fig:viewer_aif}
\end{figure}

The first viewer displays the preprocessed perfusion volumes in their original 4D form as well as maximum or minimum intensity projections, depending on the modality. Users can scroll through both spatial and temporal frames (example in Figure~\ref{fig:viewer_input}). The second viewer shows the generated 3D brain mask together with the brain segmentation, with spatial and temporal scrolling enabled (example in Figure~\ref{fig:viewer_mask}). Another viewer shows the normalized mean signal intensity over time along with the detected bolus arrival index S0 (example in Figure~\ref{fig:viewer_bolus}). In addition, the CTCs are visualized with both temporal and spatial scrolling (example in Figure~\ref{fig:viewer_ctc}). Finally, the AIF segmentation is displayed as an overlay on the CTC image. This viewer also shows the mean CTC within the AIF region and its fitted gamma variate model, again with temporal and spatial scrolling supported (example in Figure~\ref{fig:viewer_aif}).

\subsection{Technical Details}
PyPeT was developed using Python 3.10 and relies on widely used packages such as NumPy, SciPy, scikit, and matplotlib (see requirements.txt on our GitHub repository). The code was developed on Windows 11 but should run on most systems without requiring a GPU or high-performance CPU. The GitHub repository includes one demo CTP image with dimensions 200 × 200 × 23 × 45 (in-plane x, in-plane y, number of slices, number of time points). Processing this image takes about 17 seconds on an Intel i5 processor, of which 11 seconds are spent on registering the 45 three-dimensional volumes.

%% file: text/3.Experiments.tex
The performance of PyPeT was validated by comparing its output perfusion maps against reference maps obtained from established commercial and research pipelines.
PyPeT-derived perfusion maps (CBF, CBV, MTT, TTP, and Tmax) were compared against the reference maps both visually and quantitatively using the structural similarity index (SSIM).

\subsection{CT Perfusion Validation}
Three independent reference sources were used for CTP. These are 1) Icobrain CVA (FDA-approved, commercially available): ten CTP scans were selected from the publicly available ISLES2024 dataset~\cite{Riedel2024}. A dataset originally created for the Ischemic Stroke Lesion Segmentation Challenge at MICCAI 2024, which contains perfusion maps generated with Icobrain CVA; 2) iSchemaView Rapid (FDA-approved, commercially available): ten CTP scans were selected from the ISLES2018 dataset~\cite{Hakim2021,Cereda2016}. This dataset is also publicly available and was developed for stroke lesion segmentation challenges; 3) UniToBrain (Published in-house research pipeline): ten CTP scans were selected from the UniToBrain dataset~\cite{Perlo2022}. This dataset was released by the University of Turin and is also publicly accessible. In this dataset, perfusion maps were generated using an in-house pipeline based on the boxNLR model, as described in~\cite{Bennink2016}.

\subsection{MR Perfusion Validation}
One independent reference source was used for MRP, which is Olea Sphere 3.0 (FDA-approved, commercially available). Eight MRP scans were selected from an internal dataset at Inselspital, Bern, Switzerland. This dataset contains stroke patients who received DSC-MRI imaging for diagnostic purposes.

%% file: text/4.Results.tex
\begin{figure}[ht]
\centering
\includegraphics[clip, trim=1.5cm 1cm 1.5cm 0cm, width=\textwidth]{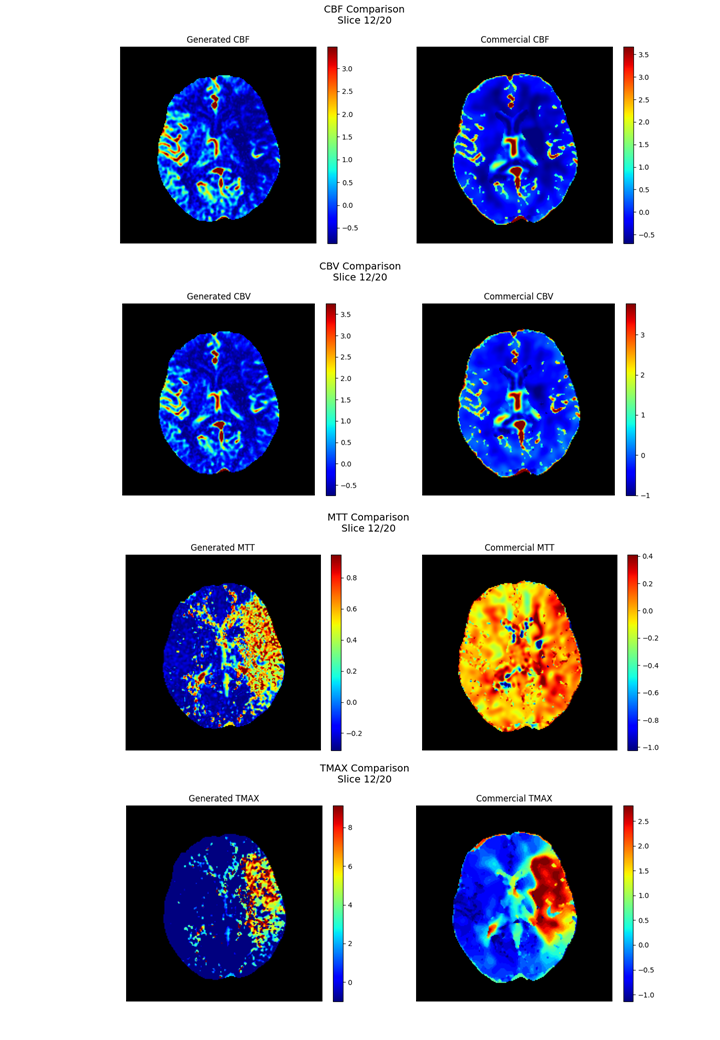}
\caption{Visual comparisons of perfusion maps generated by PyPeT with those by Icobrain CVA. This scan shown is provided as a demo image in our GitHub repository.}
\label{fig:perf_comparisons}
\end{figure}

Figure~\ref{fig:perf_comparisons} presents visual comparisons between the perfusion maps generated by PyPeT and \textit{Icobrain CVA}. Note that all maps were normalized by the whole-brain mean to obtain relative perfusion values, since deriving absolute values is challenging and absolute values as well as reported units vary across commercial software packages~\cite{Korfiatis2016,Kudo2017}. The perfusion maps generated by PyPeT show a strong correspondence with those from \textit{Icobrain CVA}. Additional visual comparisons with the other reference maps can be seen in the appendices.

A notable observation concerns the MTT maps. The PyPeT-derived MTT appears different from the corresponding \textit{Icobrain CVA} map. This discrepancy becomes clearer when examining the color bar ranges. Although the overall spatial patterns are similar, differences in color scaling create the impression of a larger divergence than is actually present. This is supported by the SSIM values reported in Table~\ref{tab:perf_comparisons}, which confirm high structural similarity between the two maps.

Table~\ref{tab:perf_comparisons} summarizes the mean SSIM and corresponding standard deviations across datasets. The results are consistent, with values typically ranging between 0.75 and 0.85. The main exception is the MTT comparison for \textit{Icobrain CVA}, which shows a lower agreement. Overall, the standard deviations are rather small, varying between 0.01 and 0.05.

\begin{table}[h]
\caption{SSIM of the comparisons between PyPeT and the reference maps.}
\label{tab:perf_comparisons}
\centering
\setlength{\tabcolsep}{12pt} % increase column spacing (default: 6pt)
\begin{tabular}{l l l c c}
\textbf{Reference}& \textbf{Modality}& \textbf{Parameter}&\textbf{Mean}& \textbf{SD}\\
\hline
Icobrain CVA& CTP&  CBF&0.83& 0.01\\
 & &  CBV&0.78&0.03\\
 & & MTT& 0.66&0.03\\
 & & Tmax& 0.76&0.04\\
 \hline
 iSchemaView Rapid& CTP& CBF& 0.83&0.05\\
 & & CBV& 0.80&0.04\\
 & & MTT& 0.75&0.04\\
& &  Tmax&0.80& 0.05\\
\hline
UniToBrain& CTP&  CBF&0.81& 0.03\\
& &  CBV&0.85& 0.02\\
 & & MTT& 0.78&0.03\\
 & & TTP& 0.80&0.02\\
 \hline
 Olea Sphere 3.0& MRP& CBF& 0.80&0.02\\
 & & CBV& 0.79&0.02\\
 & & TTP& 0.78&0.02\\
 & & Tmax& 0.80&0.02\\ 
\hline
\end{tabular}
\end{table}

%% file: text/5.Conclusion_and_Discussion.tex
This work introduced PyPeT, an open Python-based tool for generating quantitative perfusion maps from head CTP and MRP images. The tool is designed with a modular architecture, clear documentation, and low computational requirements, aiming to make perfusion imaging research as accessible and customizable as possible. PyPeT provides an extensive debug mode to facilitate transparency in the processing steps. Validation against reference perfusion maps from four different sources demonstrated consistent structural similarity, with SSIM values typically ranging between 0.75 and 0.85. While these results indicate strong correspondence with FDA approved alternatives, PyPeT is intended solely for research use. It has not undergone the level of validation required for clinical deployment.

A major challenge in validation arises from the unknown post-processing applied by commercial software, which limits the degree of visual similarity achievable between PyPeT-generated maps and reference outputs.
It is important to note that discrepancies with commercial software do not necessarily reflect inaccuracies in PyPeT-generated maps. In contrast, PyPeT offers fully transparent and reproducible results, allowing researchers to interpret and adjust the outputs more effectively than is possible with commercial tools.

It is worth noting that PyPeT remains under active development. Future work should focus on expanding functionality, improving validation strategies, and incorporating user feedback to strengthen its utility as a research resource. Some concrete points for improvement include 1) \textbf{improving absolute perfusion values:} the comparisons presented in this work were based on relative perfusion maps. To ensure that perfusion maps have accurate absolute values, it is important to tune physiological constants such as the tissue density and the hematocrit correction factor. Secondly, one may revisit the quality of arterial input function selection, and the regularization of the deconvolution process; 2) \textbf{investigation of MTT discrepancies:} although CBV and CBF maps closely match those produced by the commercial tool, the MTT maps show greater variation, especially for the comparison with Icobrain. Since MTT is theoretically the ratio of CBV to CBF, further investigation is needed to understand the post-processing strategies employed by Icobrain and to reconcile these differences; 3) \textbf{additional validation:} by applying PyPeT to additional MRP datasets, we could strengthen confidence in its generalizability and performance, especially for MRP data.

In conclusion, PyPeT lowers the barrier to entry for brain perfusion research and contributes to the growing movement toward reproducible and openly accessible medical imaging analysis.
\newline
\newline
\newline
\textbf{Acknowledgment.} This work is supported by the MIMIC project, funded by the NWO NGF AiNed XS Europe grant (NGF.1609.242.047). We also thank the Inselspital, and Richard McKinley and Roland Wiest in particular, for their discussions and for providing the MRP dataset used for validation.
\newline
\newline
\textbf{Disclosure of Interests.} The authors have no competing interests to declare that are relevant to the content of this article.

%% file: text/6.Appendices.tex
\begin{figure}[ht]
\makebox[\textwidth][c]{%
    \includegraphics[clip, trim=1.5cm 0cm 1.5cm 0cm, width=\textwidth]{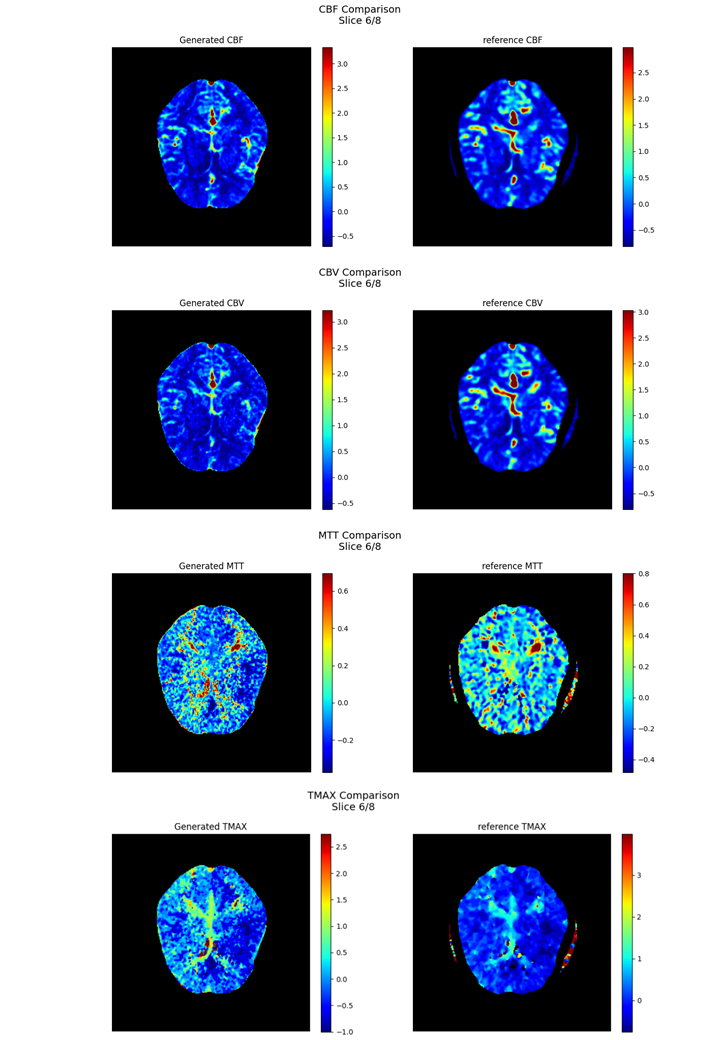}
}
\caption{Visual comparisons of perfusion maps generated by PyPeT compared to maps generated by iSchemaView Rapid.}
\label{fig:isles18_comparisons}
\end{figure}

\begin{figure}[ht]
\makebox[\textwidth][c]{%
    \includegraphics[clip, trim=1.5cm 0cm 1.5cm 0cm, width=\textwidth]{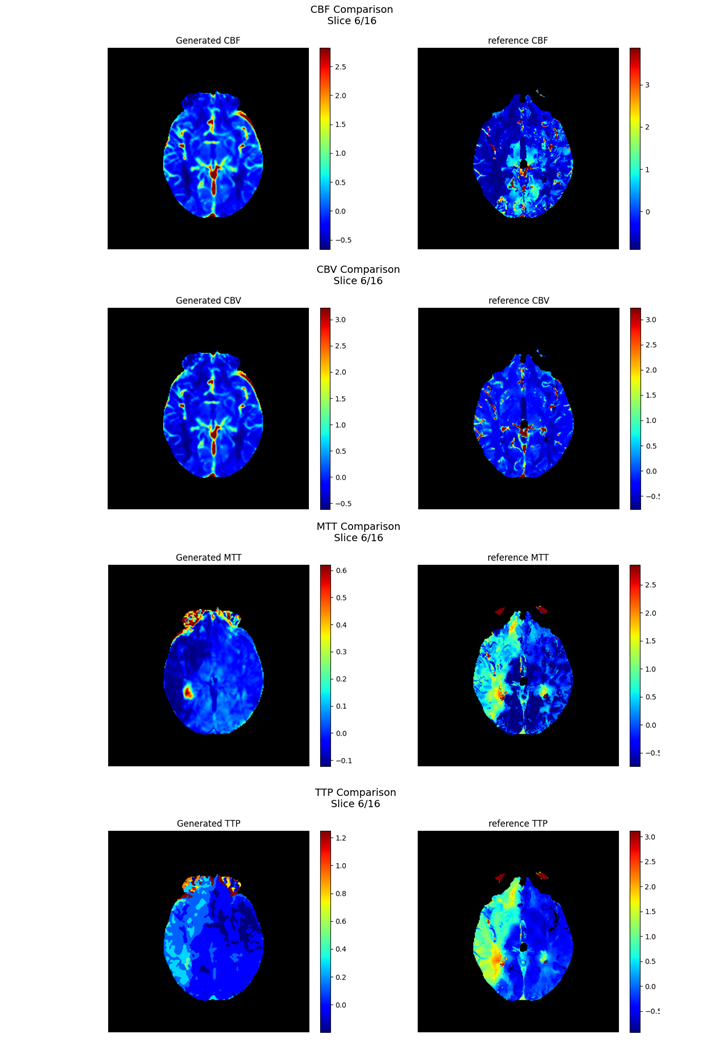}
}
\caption{Visual comparisons of perfusion maps generated by PyPeT compared to maps generated by UniToBrain.}
\label{fig:isles24_comparisons}
\end{figure}

\begin{figure}[ht]
\makebox[\textwidth][c]{%
    \includegraphics[clip, trim=1.5cm 0cm 1.5cm 0cm, width=\textwidth]{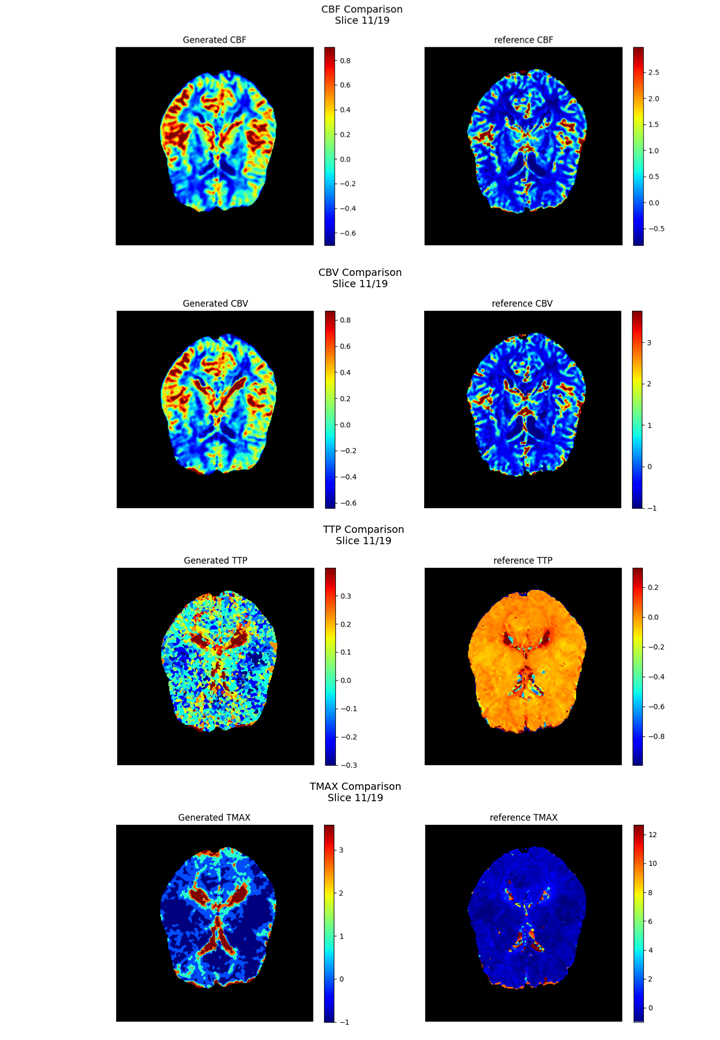}
}
\caption{Visual comparisons of perfusion maps generated by PyPeT compared to maps generated by Olea Sphere 3.0.}
\label{fig:olea_comparisons}
\end{figure}